\documentclass[fleqn,usenatbib]{mnras}
\usepackage{newtxtext,newtxmath}

\usepackage[T1]{fontenc}

\usepackage{graphicx}	
\usepackage{amsmath}	
\usepackage{amssymb}	

\newcommand{\mfp}{\lambda}

\title[Smooth MFP in semi-numerical simulations]{Improved treatments of the ionizing photon mean free path in semi-numerical simulations of reionization}

\author[F. B. Davies \& S. R. Furlanetto]{
Frederick B. Davies,$^{1,2,3}$\thanks{E-mail: davies@mpia.de (FBD)}
and Steven R. Furlanetto$^{4}$
\\
$^{1}$Department of Physics, University of California, Santa Barbara, CA 93106-9530, USA\\
$^{2}$Lawrence Berkeley National Laboratory, 1 Cyclotron Rd, Berkeley, CA 94720, USA\\
$^{3}$Max-Planck-Institut f\"{u}r Astronomie, K\"{o}nigstuhl 17, 69117 Heidelberg, Germany\\
$^{4}$Department of Physics \& Astronomy, University of California, Los Angeles, CA 90095, USA\\
}

\date{Accepted XXX. Received YYY; in original form ZZZ}

\pubyear{2022}

\begin{document}
\label{firstpage}
\pagerange{\pageref{firstpage}--\pageref{lastpage}}
\maketitle

\begin{abstract}
Efficient and accurate simulations of the reionization epoch are crucial to exploring the vast uncharted parameter space that will soon be constrained by measurements of the 21 cm power spectrum. One of these parameters, $R_{\rm max}$, is meant to characterize the absorption of photons by residual neutral gas inside of ionized regions, but has historically been implemented in a very simplistic fashion acting only as a maximum distance that ionizing photons can travel. We leverage the correspondence between excursion set methods and the integrated flux from ionizing sources to define two physically-motivated prescriptions of the mean free path of ionizing photons that smoothly attenuate the contribution from distant sources. Implementation of these methods in semi-numerical reionization codes requires only modest additional computational effort due to the fact that spatial filtering is still performed on scales larger than the characteristic absorption distance. We find that our smoothly-defined mean free path prescriptions more effectively suppress large-scale structures in the ionization field in semi-numerical reionization simulations compared to the standard $R_{\rm max}$ approach, and the magnitude of the mean free path modulates the power spectrum in a much smoother manner. We show that this suppression of large-scale power is significant enough to be relevant for upcoming 21 cm power spectrum observations. Finally, we show that in our model the mean free path plays a larger role in regulating the reionization history than in models using $R_{\rm max}$.
\end{abstract}


\section{Introduction}

Understanding the epoch of reionization, when the first stars and galaxies in the Universe ionized the hydrogen and helium in the intergalactic medium, is a key priority in observational cosmology today by virtue of its inherent connection between large-scale cosmological structure and early Universe astrophysics. The observable poised to provide the most detailed constraints on reionization is the power spectrum of the 21 cm hyper-fine structure line of neutral hydrogen (e.g. \citealt{Madau97,Furlanetto06,MW10}), which radio interferometers such as the Hydrogen Epoch of Reionization Array (HERA; \citealt{DeBoer17}), the Murchison Widefield Array (MWA; \citealt{Tingay13}), and the LOw Frequency ARray (LOFAR; \citealt{vanHaarlem13}) are aiming to detect in the near future. Indeed, substantial progress has recently been made towards mitigating systematic uncertainties in the analysis, leading to progressively lower upper limits on the 21 cm power spectrum (e.g. \citealt{Trott20,Mertens20}). Interpretation of such measurements, however, requires a detailed theoretical understanding of the patchy reionization process \citep{Furlanetto04}.

Modeling the evolving topology of reionization is an extremely challenging problem due to the large dynamic range required to probe all of the relevant physical scales. Features in the ionization field are evident up to scales of $\sim100$ Mpc, while the dominant sources of ionizing photons are likely to be hosted by dark matter halos with virial radii $r_{\rm vir}\la1$ kpc. While enormous progress has been made in high-resolution large-volume cosmological simulations of reionization (e.g. \citealt{Iliev14,Gnedin14,Dixon16,Ocvirk16,Ocvirk20,Doussot19,Trebitsch20,Garaldi21}), such undertakings are computationally expensive, and thus do not allow much freedom to explore variations in model parameters. So-called ``semi-numerical'' simulations (e.g. \citealt{MF07,Thomas09,Choudhury09,Santos10,Zahn11,Mesinger11,Majumdar14,Hutter18}), bypass the need for (hydro-)dynamical or radiative transfer calculations, instead relying on linear theory (often augmented by the Zel'dovich approximation; \citealt{Zel'dovich70}) and the excursion set formalism to construct reasonably accurate ionization fields with minimal computational expense. Such ``cheap'' simulations will allow for statistical parameter inference from future measurements of the reionization-epoch 21 cm power spectrum (e.g. \citealt{GM15,GM17,GM18,Greig20}).

One of the model parameters in the original \texttt{21cmFAST} code \citep{Mesinger11} and some other semi-numerical reionization codes (e.g. \citealt{Zahn11,AA12,Majumdar14}) is $R_{\rm max}$, an imposed upper limit to the excursion set filtering scale meant to characterize the attenuation of photons inside of reionized regions\footnote{Alternative prescriptions have also been employed which take into account integrated (e.g. \citealt{Choudhury09,Hutter18}) or instantaneous (e.g. \citealt{Hassan16}) recombination rates, or which employ an absorption-regulated ionizing background to clear out neutral islands at the end of reionization \citep{Xu17,Wu21Island}.} due to residual neutral hydrogen \citep{FO05}. But this only crudely describes absorption within ionized gas: while the attenuation along a given line of sight may be well-approximated by a step function at the location of an optically thick absorption system in the ionized gas, the attenuation when averaged over all lines of sight from a source will instead be exponential with distance, i.e. $\propto e^{-r/\mfp}$ where $\mfp$ is the mean free path (MFP) of ionizing photons. Thus the $R_{\rm max}$ prescription both underestimates the effect of absorption on scales smaller than $\mfp$, and over-suppresses the photons' ability to travel further than $\mfp$. In recent works using the Markov chain Monte Carlo reionization parameter inference code \texttt{21CMMC} \citep{GM15} (e.g. \citealt{Park19}), the constant $R_{\rm max}$ parameter has been eschewed entirely in favor of the inhomogeneous recombination prescription from \citet{SM14}. However, the underlying method by which the absorption is implemented is effectively a spatially variable $R_{\rm max}$. In addition, while the prescription from \citet{SM14} takes into account additional physics from the history and spatial variations in the recombination rate, it relies on a particular sub-grid model for density fluctuations \citep{MHR00} -- indeed, the physics of the mean free path may be far more complicated due to the relaxation of clumpy gas following photoheating by ionization fronts \citep{Park16,D'Aloisio20}. Inhomogeneous recombinations also do not explicitly account for the \emph{distance} between sources of ionizing photons and the regions being ionized. The mean free path offers the advantage of physical transparency over these more detailed approaches, matches more smoothly onto descriptions of the ionizing background after reionization, and it can be more easily included in the parameter inference process. The development of a robust and accurate mean free path scheme in semi-numerical simulations is therefore of crucial importance for interpreting the science from future observations.

In this work, we present two new methods for implementing the mean free path of ionizing photons in semi-numerical simulations of reionization topology. Our new methods are motivated by a direct correspondence between how the excursion set method is employed by \texttt{21cmFAST} -- i.e., via central pixel flagging -- and an accounting of the number of ionizing photons that reach any given point from surrounding sources. The first method is a straightforward adjustment of the ionization criterion that can be applied to any semi-numerical model, while the second method uses a modified excursion set filter shape which is only suitable for semi-numerical models with discrete sources. To facilitate use of the latter method in halo-free simulations, we also introduce an ionization field prescription (``FFRT-P'') that treats ionizing sources as discrete without explicitly forming or accounting for individual dark matter halos.

The rest of the paper is structured as follows. In Section~\ref{sec:count}, we describe how excursion set methods are connected to simple photon counting. In Section~\ref{sec:mfp} we use this counting analogy to derive our smooth mean free path implementations and show how they affect the resulting structure of the ionization field. In Section~\ref{sec:obs} we show the implications for future observations of the 21 cm power spectrum. We then conclude in Section~\ref{sec:conc}.

In this work we assume cosmological parameters from \emph{Planck} with $(h,\Omega_m,\Omega_\Lambda,\Omega_b,\sigma_8,n_s) = (0.6736, 0.3153, 0.6847, 0.0493, 0.8111, 0.9649)$ \citep{Planck18}. Distance units are comoving unless specified otherwise.

\section{Photon Counting in Semi-Numerical Simulations} \label{sec:count}

Semi-numerical simulations of reionization were inspired by a simple ``photon-counting" argument originally presented by \citet{Furlanetto04}. If we assume that all of the ionizing photons generated in a region are absorbed within that region, then the condition for the region to be ionized is simply $\zeta f_{\rm coll}\geq1$, where $\zeta$ is the number of ionizing photons emitted\footnote{Note that we neglect the contribution of recombinations in our definition of $\zeta$ to avoid double-counting them when accounting for absorption.} per collapsed mass (also known as the ionizing efficiency) and $f_{\rm coll}$ is the fraction of mass in collapsed objects above some minimum halo mass $M_{\rm min}$. The excursion set formalism can then be used to find the maximum scale for which the criterion is satisfied, which the analytic model then identifies as an ionized region. The condition on the collapsed fraction is often referred to as ``photon counting" because it depends on the cumulative production of photons over cosmic history. Unfortunately, that approach makes it difficult to incorporate photons that are absorbed by neutral gas into the criterion (aside from in an average sense). 

In the excursion set formalism, when a region reaches the critical threshold for collapse, the entirety of the mass (i.e., the region in Lagrangian space) is then contained in the collapsed structure \citep{Bond91}. \citet{MF07} implemented this concept in a semi-numerical reionization code by ionizing entire spherical regions corresponding to the first filter size (starting from the size of the simulation box and going to smaller scales) at which they crossed the ionization threshold. However, naive methods of painting these spherical regions onto three dimensional grids become quite slow when the total ionized volume is large, i.e. after the midpoint of reionization. To compute ionization fields more efficiently, \citet{Zahn11} instead chose to only ionize individual pixels of the filtered density (or halo mass) field which crossed the ionization threshold, the so-called ``central pixel flagging'' method (see also \citealt{Mesinger11}). Such methods compare favorably to full radiative transfer simulations (e.g. \citealt{Majumdar14}). As we show below, the central pixel flagging implementation is essentially identical to asking whether the \emph{time-integrated ionizing photon flux} reaching the central pixel is enough to ionize the gas -- a subtle but conceptually important shift from the standard photon-counting argument.

The fundamental idea of semi-numerical simulations is to count the number of ionizing photons produced in some volume, and if the number of ionizing photons exceeds the number of hydrogen atoms in the same volume, the region is then ionized. In the simplest case, we first consider the photons that reach a cell of volume $\Delta V = \Delta r^3$ from a neighboring region a comoving distance $r$ away. Neglecting absorption, the ionizing flux reaching the cell is
\begin{equation}
f = \frac{\dot{N}_{\rm ion}}{4\pi r_p^2},
\end{equation}
where $r_p=r/(1+z)$ is the proper distance, $\dot{N}_{\rm ion}=\epsilon\Delta V$ is the rate at which ionizing photons are produced in the neighboring region, and $\epsilon$ is the emissivity of ionizing photons. The number of photons entering the cell from the neighboring region per unit time $\dot{N}_1$ is then the flux multiplied by the surface area of the cell $\Delta r_p^2$,
\begin{equation}
\dot{N}_1 = \frac{\epsilon\Delta V}{4\pi}\,\frac{\Delta r_p^2}{r_p^2} = \frac{\epsilon\Delta V}{4\pi}\,\frac{\Delta r^2}{r^2}.
\label{eqn:n1}
\end{equation}

Now we sum the emissivity from all regions at a constant comoving distance $r$ from the cell of interest, located in a shell of radius $r$ and thickness $\Delta r$. The shell's comoving volume is $4 \pi r^2 \Delta r$, so in equation~(\ref{eqn:n1}) the total number of incident photons per unit time emitted from these sources is
\begin{equation}
\dot{N}_{\rm sh} = \epsilon \Delta r^3.
\end{equation}
If we now integrate this ionizing photon emission over the entire history of the Universe, we have
\begin{equation} \label{eqn:shell}
N_{\rm sh} = \zeta f_{\rm coll} n_b^{\rm sh} \Delta r^3,
\end{equation}
where $\zeta$ is the number of ionizing photons produced per baryon in dark matter halos, $n_b^{\rm sh}$ is the baryon density inside the shell, and $f_{\rm coll}$ is the fraction of baryons inside dark matter halos within the shell.\footnote{Note that, for simplicity, we have assumed here that the ionizing efficiency is time-independent.} The total number of baryons within the cell under consideration is $n_b^0\Delta r^3$, so the cell will be ionized if 
\begin{equation} \label{eqn:shell2}
\zeta f_{\rm coll} > \frac{n_b^0}{n_b^{\rm sh}}.
\end{equation}

Furthermore, the steps leading to equation~(\ref{eqn:shell}) show that the contribution to the ionizing photon budget from each radial shell is constant. We can therefore make the simple ansatz that, on average, spherical symmetry is a good approximation. Then we can replace the contribution from each shell with the average contribution over an entire spherical region. If we generalize our cell to be that entire region, so $n_b^0=n_b^{\rm sh}$, we then recover the standard ionization criterion \citep{Furlanetto04},
\begin{equation}
\zeta f_{\rm coll} > 1.
\end{equation}
This is identical to the standard photon-counting argument.

In summary, the standard central-pixel-flagging semi-numerical method for approximating reionization topology is more or less identical to simply integrating the \emph{ionizing flux} from surrounding sources. In the next section, we show how this analogy allows us to implement the mean free path of ionizing photons in a novel way.

\begin{figure}
\begin{center}
\resizebox{8.5cm}{!}{\includegraphics[trim={1em 1em 1em 1em},clip]{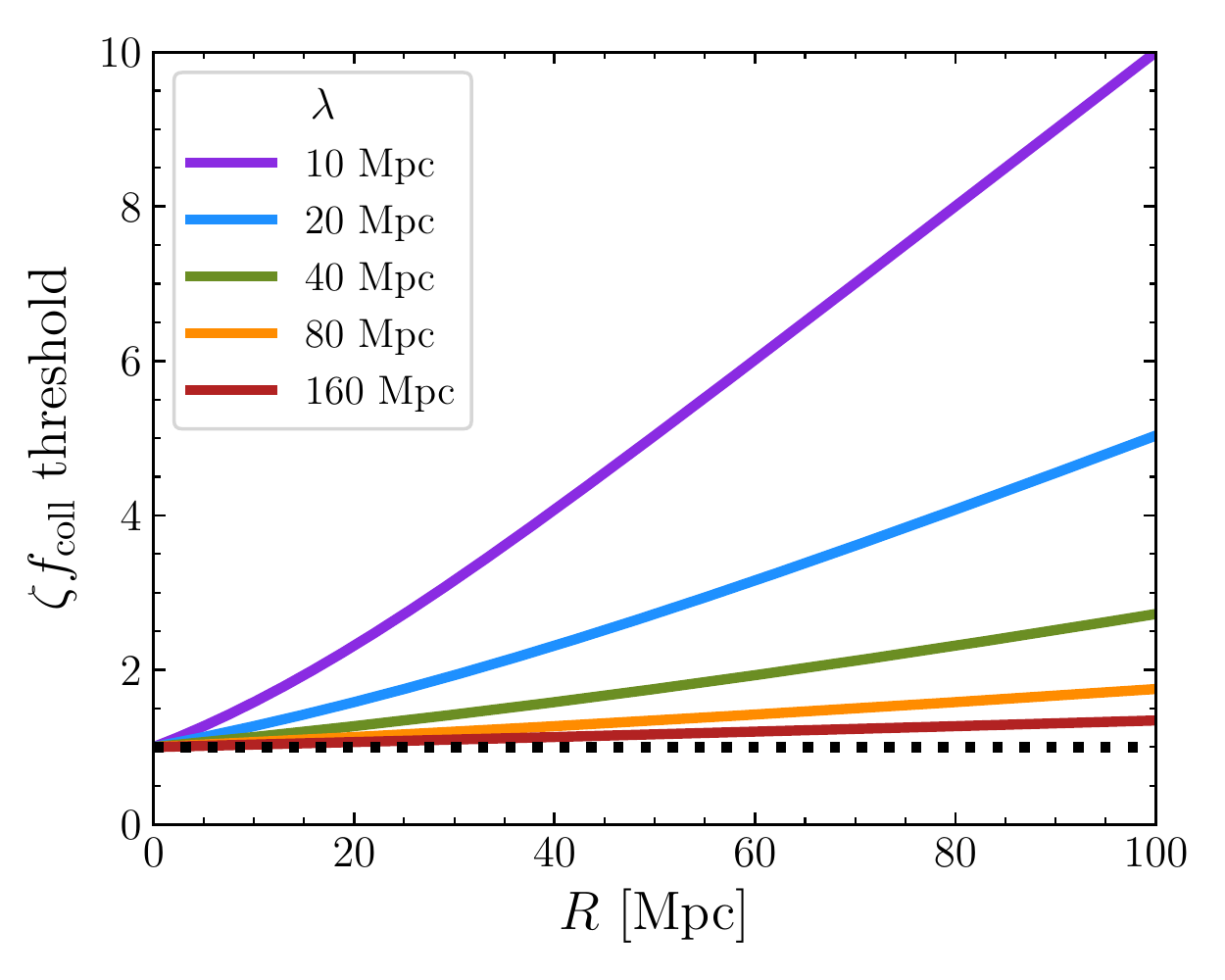}}
\end{center}
\caption{Scale-dependent ionization thresholds (equation~\ref{eqn:newcrit}) for $\mfp=$ 10, 20, 40, 80, and 160 Mpc from top to bottom (solid curves) compared to the standard static threshold (dotted line). The required $\zeta f_{\rm coll}$ increases roughly linearly with $R$ when $R>\mfp$.}
\label{fig:mfpcrit}
\end{figure}

\section{New Semi-Numerical Implementations of the Mean Free Path} \label{sec:mfp}

In the previous section, we neglected the absorption of ionizing photons by residual neutral hydrogen in the ionized medium as those photons travel through the ionized gas to the central cell. This absorption attenuates the ionizing flux by a factor of $e^{-r/\mfp}$ on average, where $\mfp$ is the mean free path of ionizing photons. Equation~(\ref{eqn:shell}) then becomes
\begin{equation}
N_{\rm sh} = \zeta f_{\rm coll} n_b^{\rm sh} \Delta r^3 e^{-r/\mfp}.
\end{equation}
Unlike the original derivation, this criterion does depend on the shell's distance from the central cell. Nevertheless, assuming that the counting argument described in the previous section holds, we can estimate the effect of attenuation by comparing the integrated number of ionizing photons from spherical shells of width $dr$ that reaches the central cell with and without attenuation. If $N_{\rm tot}$ is the total number of ionizing photons reaching the central cell out to a distance $R$ in the absence of absorption, while $N_{\rm tot}^\mfp$ is the total number with absorption, this comparison yields
\begin{equation} \label{eqn:atten}
\frac{N_{\rm tot}}{N_{\rm tot}^{\mfp}} = \frac{\int_0^R \zeta f_{\rm coll} n_b^{\rm sh} \Delta r^2 dr}{\int_0^R \zeta f_{\rm coll} n_b^{\rm sh} \Delta r^2 e^{-r/\mfp} dr} = \frac{R}{\mfp \left[1-e^{-R/\mfp}\right]}.
\end{equation}
This ratio shows the number of ``excess" ionizing photons required in a spherical region to account for uniform absorption within that region. 
We note that while the derivation of equation~(\ref{eqn:atten}) relies on rough approximations derived in the previous section, we have confirmed that the expression is accurate in simple 3D numerical experiments of uniform emissivity spheres with uniform opacity.

From the previous expression we can then define a scale-dependent ionization criterion that accounts for the absorption of ionizing photons inside the ionized bubble,
\begin{equation} \label{eqn:newcrit}
\zeta f_{\rm coll} > \frac{R}{\mfp \left[1-e^{-R/\mfp}\right]} \qquad \qquad \mbox{(MFP-}\bar{\epsilon} \ \mbox{method})
\end{equation}
This equation provides a revised ionization threshold for semi-numerical schemes, with the key assumptions of uniform absorption, uniform emissivity, and a constant $\mfp$. In particular, the derivation assumes that sources of ionizing photons are uniformly distributed inside of the radius $R$, i.e. that the ionizing emissivity $\epsilon$ (or equivalently, the halo mass density) averaged on spherical shells is independent of radius. As shown in Figure~\ref{fig:mfpcrit}, when $R\ll\mfp$ we recover the standard criterion, while for $R>\mfp$ the cell becomes more and more difficult to ionize, as expected.  In the following, we will refer to this implementation of the MFP as the ``MFP-$\bar{\epsilon}$'' method.

In practice, however, the radial distribution of sources around a given cell will be non-uniform, and this inhomogeneity will affect how many ionizing photons are absorbed along the way. For example, if there is only one source of ionizing photons, the degree of attenuation will depend on its distance from the center. In this more general case, the integral in the denominator of equation~(\ref{eqn:atten}) does not have a simple analytic solution, and must be instead explicitly computed at every location. 

\begin{figure}
\begin{center}
\resizebox{8.5cm}{!}{\includegraphics[trim={1em 1em 1em 1em},clip]{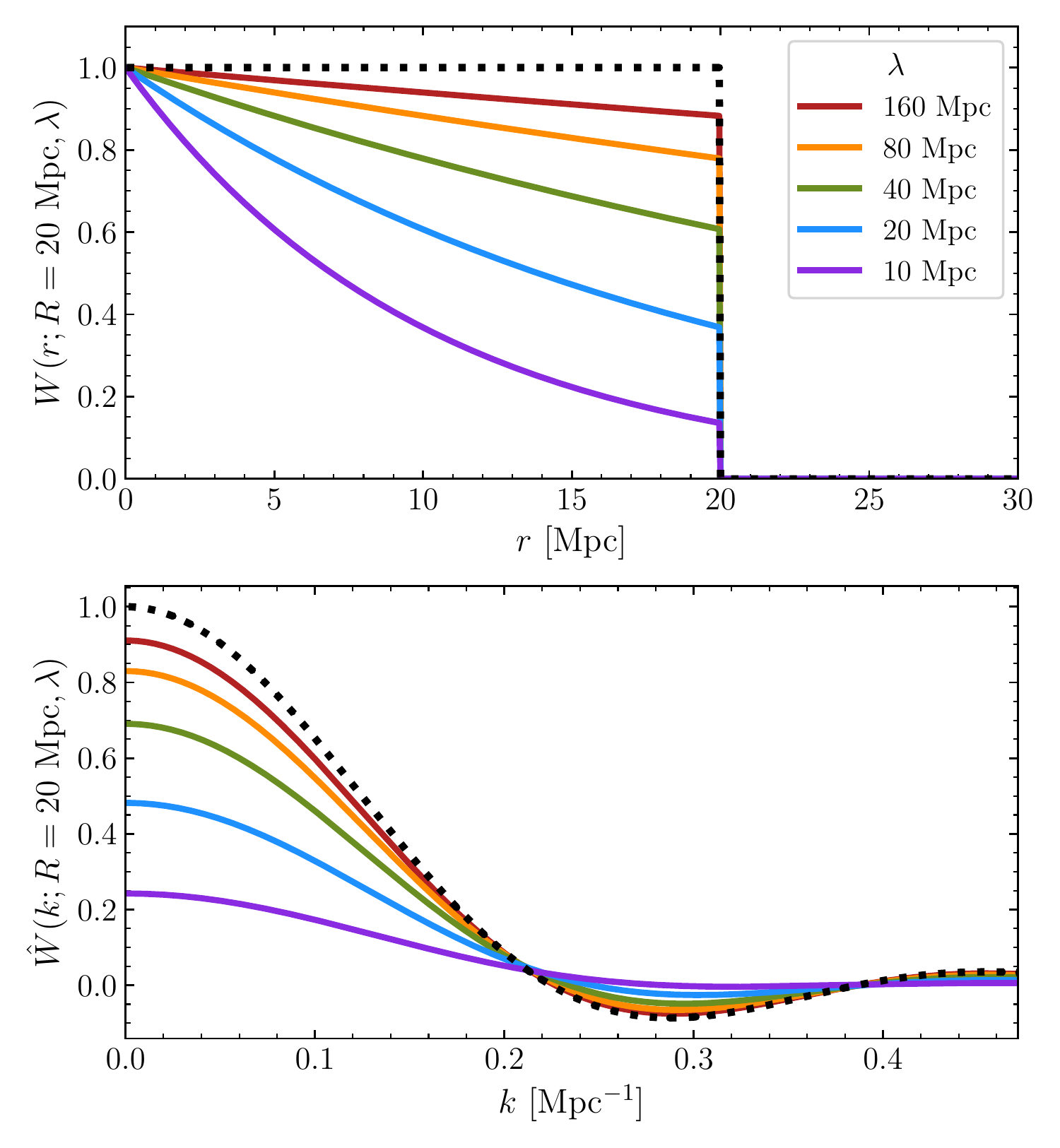}}
\end{center}
\caption{Exponential spherical top-hat filters in real space ($W(r;R,\mfp)$, top panel) and in $k$-space ($\hat{W}(k;R,\mfp)$, bottom panel) for $\mfp$\,=\,10, 20, 40, 80, and 160 Mpc from top to bottom (solid curves) compared to the standard spherical top-hat filter (dotted line), assuming a filter scale $R=20$\,Mpc.}
\label{fig:mfpfilter}
\end{figure} 

Fortunately, if we think of the semi-numeric procedure to calculate the ionization field, we can see how to approach this problem in a straightforward manner. At each step in the standard semi-numerical process, the density field is filtered over a spherical region subtending a scale $R$ and then converted into a number of emitted ionizing photons via $f_{\rm coll}$ and $\zeta$. The integral for the general ionizing source distribution is equivalent to a convolution of the ionizing emissivity field with a spherical top-hat filter multiplied by the factor $e^{-r/\mfp}$, i.e.
\begin{equation}
W(r; R,\mfp) = \begin{cases}
\frac{3}{4\pi R^3}  e^{-r/\lambda} &\text{if } r < R\\
0 &\text{if } r \geq R,
\end{cases}
\end{equation}
for filter scale $R$, so the integration can be efficiently computed in Fourier space. The Fourier transform of this ``exponential spherical top-hat filter'' has an analytic solution,
\begin{equation} \label{eqn:mfpfilt}
\begin{split}
\hat{W}(k;R,\mfp) &= \frac{-3\mfp}{kR^3 (k^2\mfp^2+1)^2}\\
&\times \Big(e^{-R/\mfp}\big[(k^2\mfp^2R+2\mfp+R)k\mfp\cos{(kR)} \\
&+(-k^2\mfp^3+k^2\mfp^2R+\mfp+R)\sin{(kR)}\big]-2k\mfp^2\Big),
\end{split}
\end{equation}
which is convenient (although not explicitly required) for semi-numerical calculations. In Figure~\ref{fig:mfpfilter}, the solid curves show examples of this filter at a fixed filter scale $R=20$ Mpc with varying $\mfp$ in real space (top panel) and in $k$-space (bottom panel) compared to the standard spherical top-hat, shown by the dotted curve. 

We can thus build the ionization field while explicitly incorporating the emissivity fluctuations by replacing the standard spherical top-hat filter by this modified version at each filter scale, in effect solving the following barrier, 
\begin{equation}\label{eqn:mfp-ebar}
\int_0^\infty \langle \zeta f_{\rm coll} \Delta \rangle_r W(r; R,\mfp)4\pi r^2dr > \langle \Delta(<R) \rangle,\  \mbox{(MFP-}\epsilon(r)\ \mbox{method})
\end{equation}
where $\Delta\equiv\rho/\bar{\rho}$ is the matter overdensity, $\langle \zeta f_{\rm coll} \Delta \rangle_r$ represents the average product of the three quantities in all cells within a spherical shell at radius $r$, and $\langle \Delta(<R) \rangle$ is the mean overdensity of the entire spherical volume within the filter of radius $R$. In the following we will refer to this implementation as the ``MFP-$\epsilon(r)$'' method.

While this second method should in general capture the effect of ionizing photon absorption better than the first, it requires knowledge of the \emph{spatial distribution} of sources inside the filter. The distribution of sources is not captured by the most efficient semi-numerical reionization method ``FFRT'' \citep{Zahn05,Zahn07,Zahn11}, popularized by the public code \texttt{21cmFAST} \citep{Mesinger11}, which computes $f_{\rm coll}$ directly from the filtered density field following a conditional Press-Schechter prescription \citep{LC93}. The MFP-$\epsilon(r)$ method should instead only be applied to methods that can produce the $f_{\rm coll}$ field derived from discrete halos (e.g. \texttt{DexM}, \citealt{MF07}, see also \citealt{Choudhury09,Santos10,Majumdar14,Hutter18}). However, these halo-based methods tend to have limited dynamic range due to the requirement that the initial conditions resolve the minimum halo mass (although arbitrarily low mass halos can in principle be ``painted on'' as in \citealt{McQuinn07}). To partially overcome this limitation we have developed a method -- which we dub ``FFRT-P'' -- to bridge the gap between the techniques of \texttt{21cmFAST} and \texttt{DexM}. The FFRT-P approach computes $f_{\rm coll}$ in the same way as \texttt{21cmFAST}, but only once, at the \emph{pixel scale} of the evolved density field in the semi-numerical simulation. The resulting collapsed mass field is then filtered as in, e.g., \texttt{DexM}. The detailed implementation of FFRT-P, a direct comparison to FFRT and \texttt{DexM}, and some of its limitations are described in the Appendix.

\begin{figure*}
\begin{center}
\resizebox{11.5cm}{!}{\includegraphics[trim={1em 1em 1em 1em},clip]{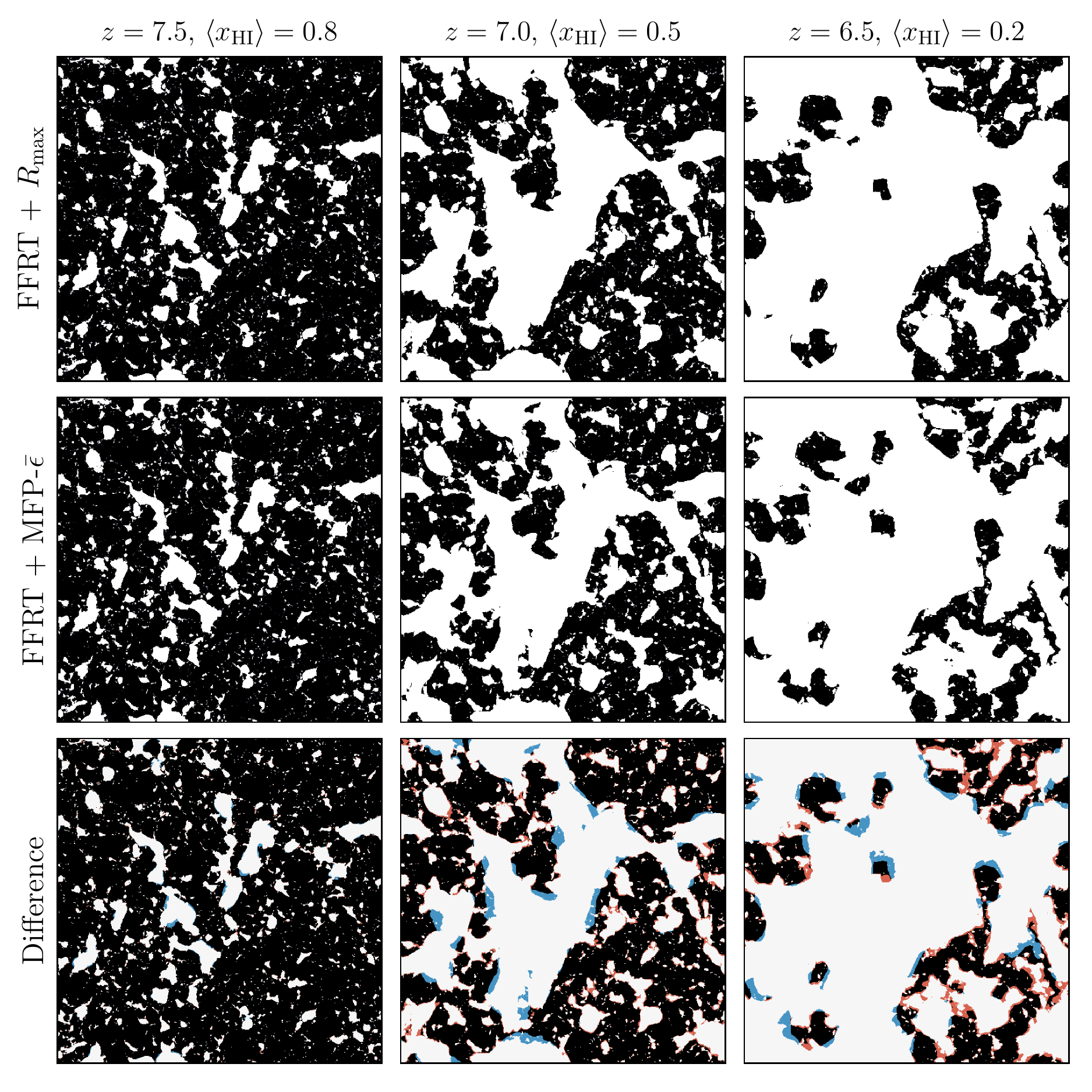}}
\end{center}
\vskip -2em
\caption{Comparison between slices of a semi-numerical reionization simulation using the standard FFRT method with a maximum filter scale $R_{\rm max}=20$ Mpc (top row) and our modified ionization criterion (MFP-$\bar{\epsilon}$) with $\lambda=20$ Mpc (second row), where black regions are neutral and white regions are ionized. The third row highlights the difference between the two slices: white and black regions are the same for both, while blue and red areas highlight regions which are only ionized in the standard FFRT and MFP-$\bar{\epsilon}$ methods, respectively. From left to right the panels show models with volume-averaged neutral fractions of $\langle x_{\rm HI}\rangle=0.8$, 0.5, and 0.2, at $z=7.5$, 7.0, and 6.5, respectively. The slices shown are 0.5 Mpc thick and 256 Mpc on a side.}
\label{fig:ffrt_slices}
\end{figure*}

\begin{figure}
\begin{center}
\resizebox{8.0cm}{!}{\includegraphics[trim={1em 1em 1em 1em},clip]{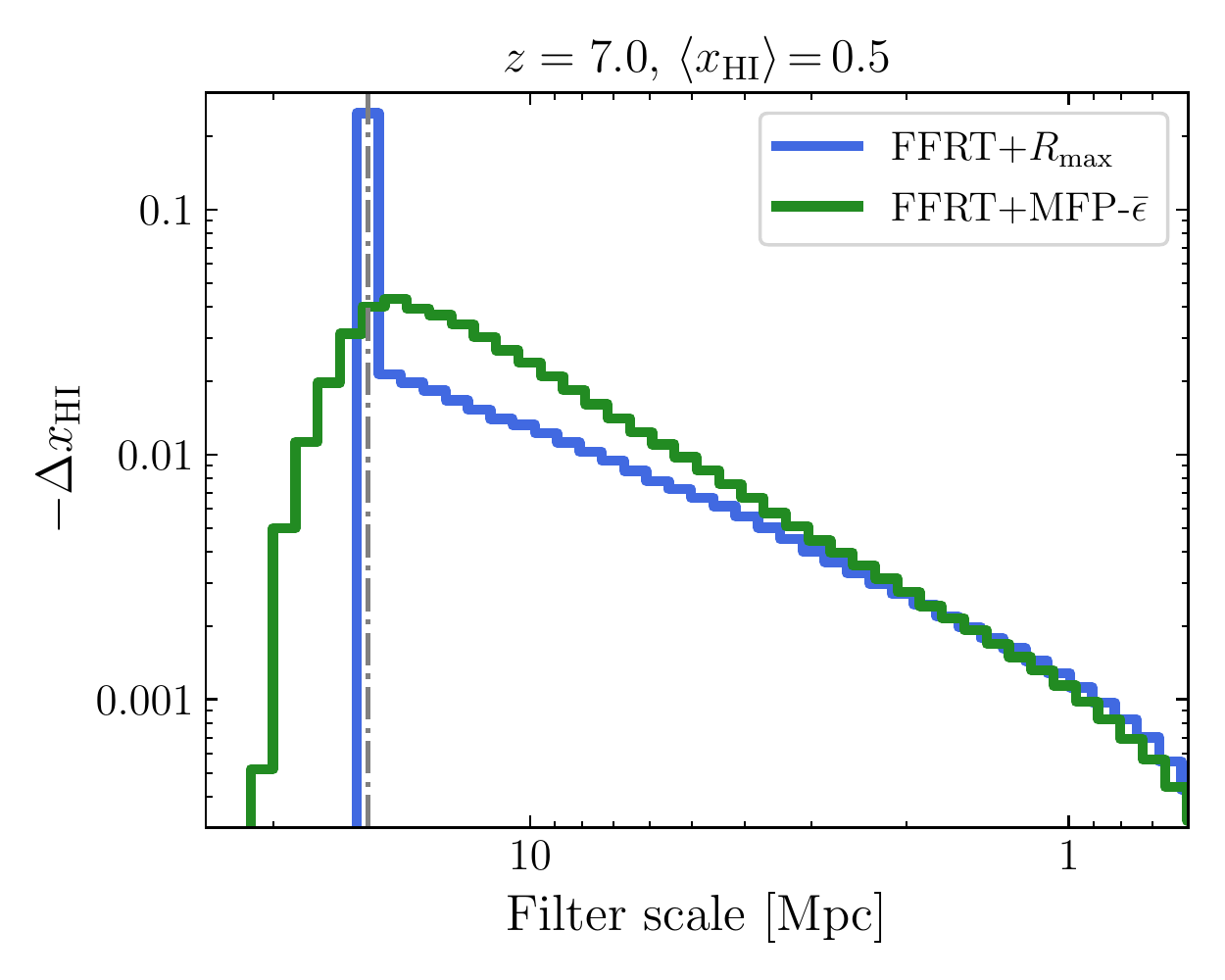}}
\end{center}
\caption{Change in neutral fraction due to each filter scale in 21cmFAST using the standard FFRT+$R_{\rm max}$ approach (blue) and our new FFRT+MFP-$\bar{\epsilon}$ method (green). The size of $R_{\rm max}$ and $\mfp$ is shown by the vertical dash-dotted line.}
\label{fig:dxdr}
\end{figure}

\subsection{21cmFAST implementation}

\begin{figure*}
\begin{center}
\resizebox{11.5cm}{!}{\includegraphics[trim={1em 1em 1em 1em},clip]{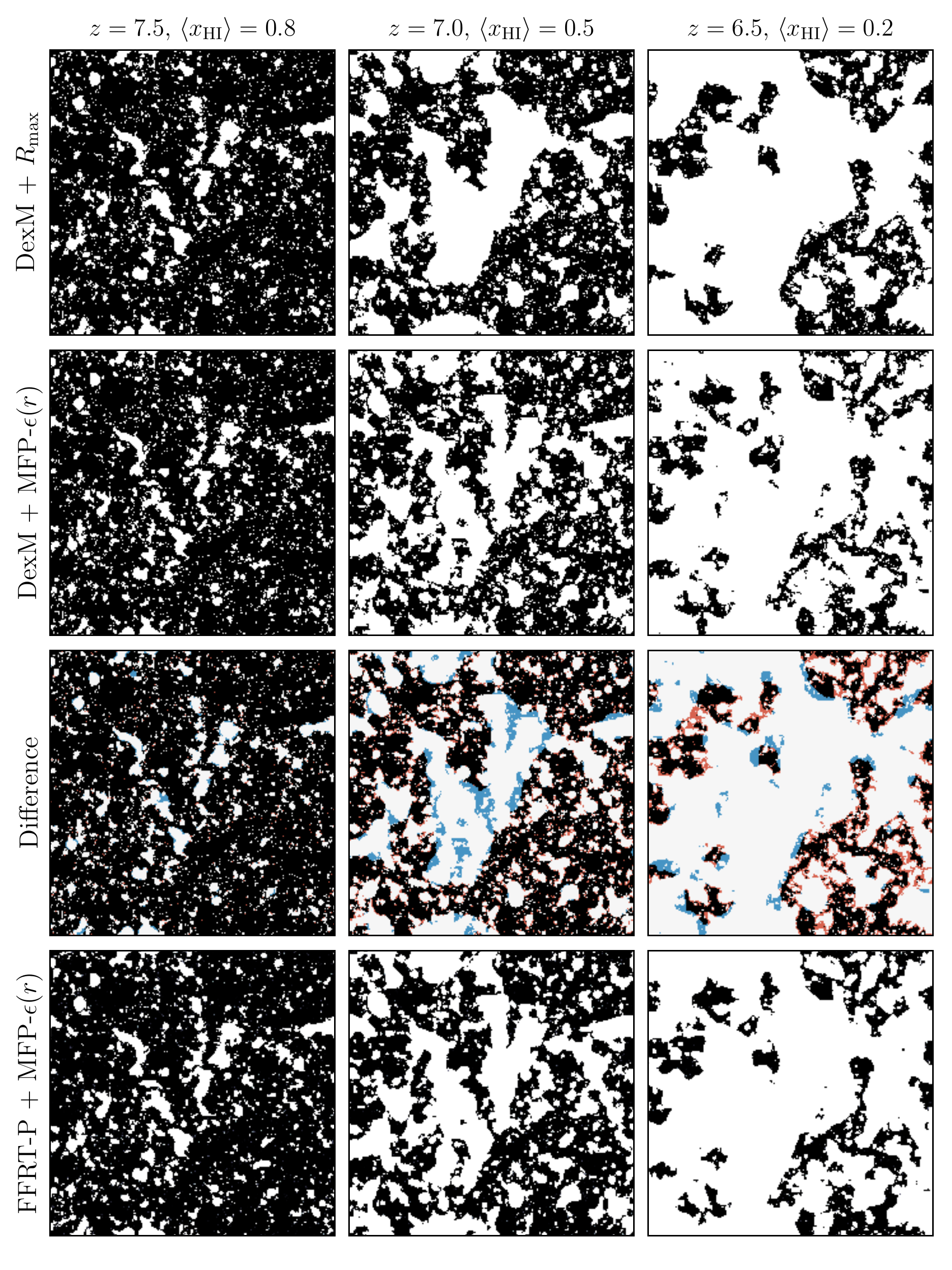}}
\end{center}
\vskip -2em
\caption{Similar to Figure~\ref{fig:ffrt_slices} but now comparing the halo-based approaches with a maximum filter scale $R_{\rm max}=20$ Mpc (top row) and the MFP-$\epsilon(r)$ method with $\lambda=20$ Mpc (second row). The third row shows the difference between the two, similar to the bottom panel of Figure~\ref{fig:ffrt_slices}. The bottom row shows our new FFRT-P method with MFP-$\epsilon(r)$ filtering ($\lambda=20$ Mpc). The slices shown are 0.5 Mpc thick and 256 Mpc on a side.}
\label{fig:dexm_slices}
\end{figure*}

We have implemented both of the mean free path methods described above by modifying the semi-numerical reionization code \texttt{21cmFAST} \citep{Mesinger11}. Here we present comparison tests from a simulation 256 Mpc on a side with $4096^3$ initial conditions and $512^3$ evolved fields (i.e. density, ionization, and 21 cm brightness temperature). We assume a minimum halo mass of $M_{\rm min}=10^9 M_\odot$ and tune the ionizing efficiency of each method to achieve volume-averaged neutral hydrogen fractions of $\langle x_{\rm HI}\rangle =$ 0.2, 0.5, and 0.8 at $z=$ 6.5, 7.0, and 7.5, respectively, following a reionization history consistent with the latest constraints from the CMB optical depth \citep{Planck18} and other recent astrophysical constraints (e.g. \citealt{Greig17b,Mason18,Davies18b}). As in \citet{DF16}, we smooth the evolved density field by oversampling the initial conditions and interpolating the displacement field when computing the Zel'dovich approximation. This smoothing procedure substantially reduces the impact of shot noise close to the pixel scale of the evolved density field, as will be evident in the power spectra presented below, but it does not otherwise change any of our primary results. For each redshift we also constructed distributions of discrete dark matter halos down to $M_{\rm min}$ from the initial conditions using the excursion set halo finder originally developed for the \texttt{DexM} code \citep{MF07}, and updated their positions using the Zel'dovich approximation displacements (see \citealt{MF07} for details). In the following we will refer to the ionization field method utilizing these halos as ``DexM,'' and we will refer to the default \texttt{21cmFAST} ionization field method, derived directly from the density field, as ``FFRT.''

\begin{figure*}
\begin{center}
\resizebox{16cm}{!}{\includegraphics[trim={1em 1em 1em 1em},clip]{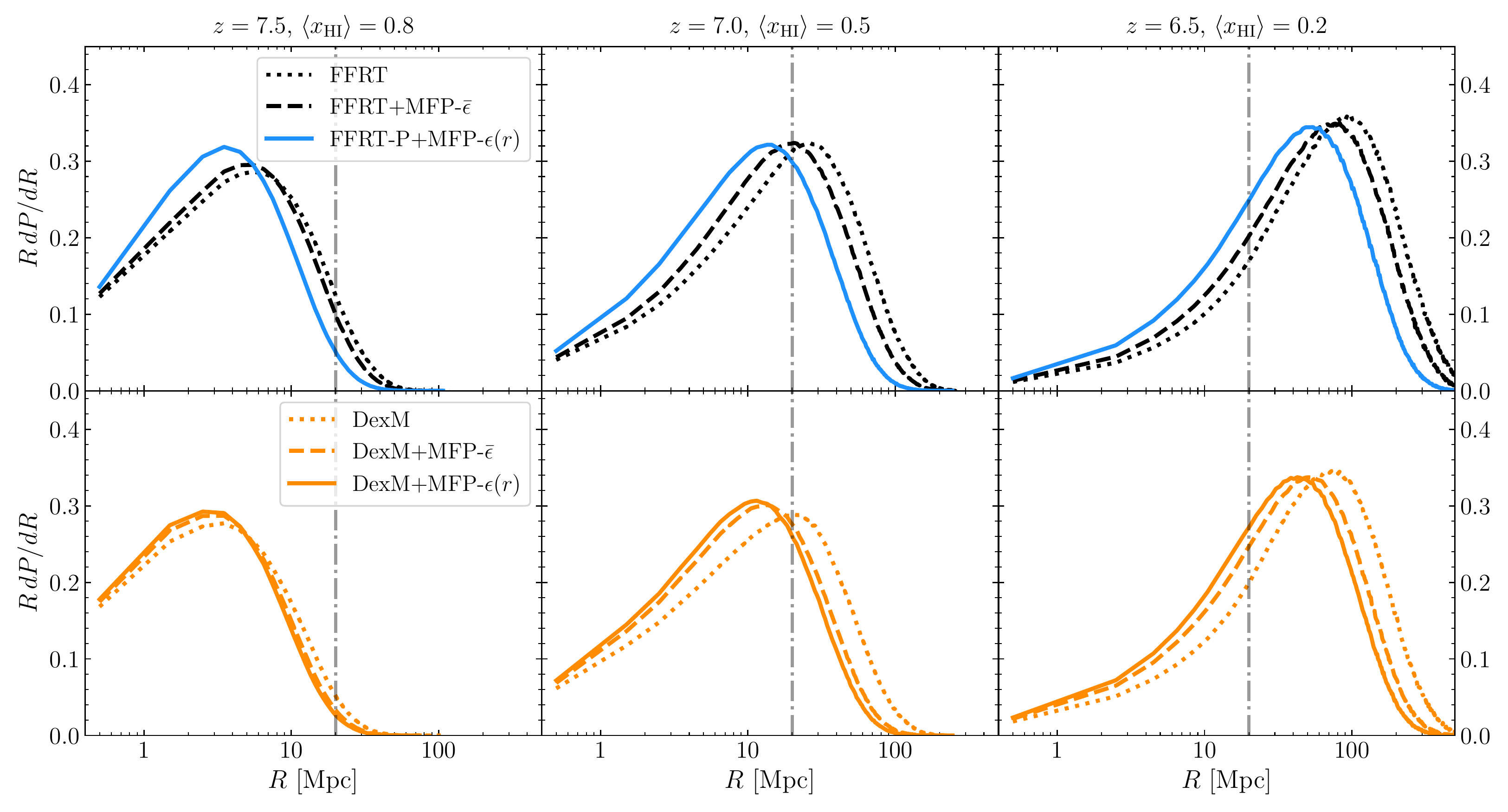}}
\end{center}
\vskip -2em
\caption{Ionized bubble sizes for $\langle x_{\rm HI}\rangle=0.8$, 0.5, and 0.2 at $z=7.5$, 7.0, and 6.5, respectively, from left to right, assuming a mean free path of 20 Mpc (shown by the vertical line). Dotted curves show models which use $R_{\rm max}$ to approximate the effect of the mean free path, while dashed and solid curves use the MFP-$\bar{\epsilon}$ and MFP-$\epsilon(r)$ methods, respectively, introduced in this work. The top panels show the ``halo-free'' methods FFRT (black) and FFRT-P (blue), while the lower panels show methods with \texttt{DexM} halos (orange). The smooth mean free path methods tend to shift the distributions to smaller sizes.}
\label{fig:sizes_redshift}
\end{figure*}

\begin{figure*}
\begin{center}
\resizebox{16cm}{!}{\includegraphics[trim={1em 1em 1em 1em},clip]{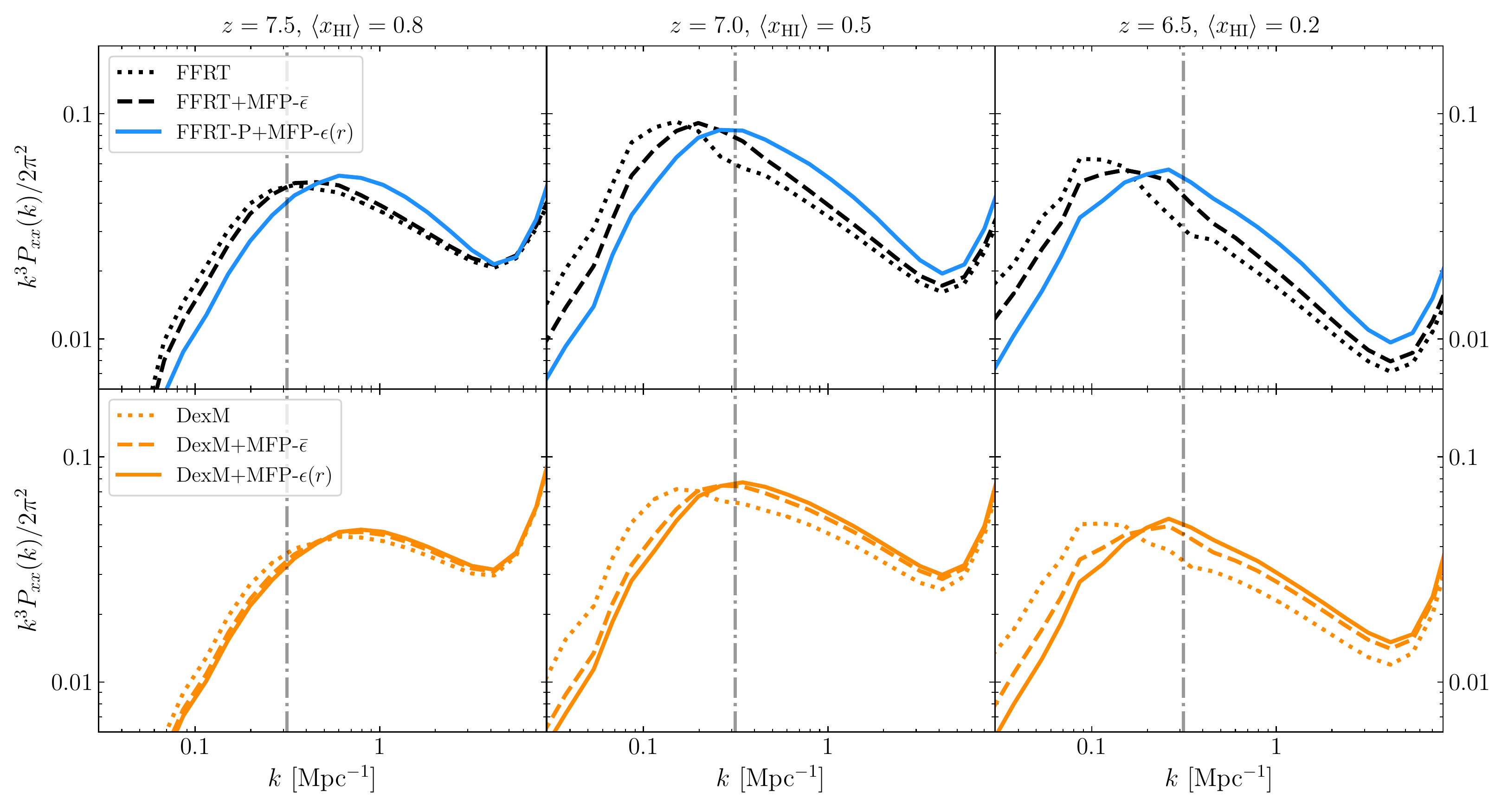}}
\end{center}
\vskip -2em
\caption{Similar to Figure~\ref{fig:sizes_redshift} but now showing the ionization field power spectra. The vertical line corresponds to the mean free path scale, $k=2\pi/\mfp$ where $\mfp=20$ Mpc. The new smooth mean free path methods tend to shift power from large to small scales at fixed $\langle x_{\rm HI} \rangle$.}
\label{fig:pxx_redshift}
\end{figure*}

In Figure~\ref{fig:ffrt_slices} we show slices from the ionization fields employing the standard FFRT approach with $R_{\rm max}=20$ Mpc (top panels), FFRT with MFP-$\bar{\epsilon}$ (equation~\ref{eqn:newcrit}) and $\lambda=20$ Mpc (middle panels), and the difference between the two (bottom panels). This mean free path is somewhat smaller than the extrapolation from observations of stacked quasar spectra\footnote{We note that the mean free path along quasar sightlines at $z\sim5$ could be biased high due to the proximity effect \citep{D'Aloisio18,Davies20ghost}, but this has not yet been detected by observations (see \citealt{Becker21}).} at $z\lesssim5$ by \citet{Worseck14}, somewhat larger than the recent mean free path measurement at $z\sim6$ by \citet{Becker21}, but consistent with the range of possible values for $z\sim7$ from the high-resolution hydrodynamical simulations of \citet{D'Aloisio20} (see also \citealt{Cain21}). At high neutral fraction when the ionized bubbles are still small compared to $R_{\rm max}$ or $\mfp$, the ionization fields are nearly identical, as expected. However, at lower neutral fractions, differences begin to arise, namely it is apparent that the sizes of large ionized regions are suppressed when the ionization threshold of the MFP-$\bar{\epsilon}$ method is employed. This may be a surprise given that the $R_{\rm max}$ prescription more harshly decreases the impact of distant photons, but the smooth attenuation of photons at scales smaller than $\mfp$ appears to have a stronger effect on the resulting ionization field. The increased size of smaller ionized regions when applying the MFP-$\bar{\epsilon}$ method, shown by the red regions in the bottom panels of Figure~\ref{fig:ffrt_slices}, is then simply a consequence of the increased $\zeta$ required to keep the global neutral fraction constant.

Recall that the application of the excursion set formalism in \texttt{21cmFAST} involves convolutions of the density field with a series of filters, defining discrete spatial scales at which the ionization barrier is tested. In Figure~\ref{fig:dxdr}, we show the sequential change in the volume-averaged neutral fraction -- i.e., the number of ionizations -- due to each excursion set filtering scale for the FFRT+$R_{\rm max}$ and FFRT+MFP-$\bar{\epsilon}$ approaches. Note that the excursion set method works from large to small scales, and hence the horizontal axis has been flipped such that the progression is from left to right. The hard cutoff scale of the standard method leads to a sharp excess in ionization at $R_{\rm max}$, whereas in our new method the effect is much smoother, as expected. The end result is that a wider range of scales provide substantial contributions to the ionization field.

Similar to Figure~\ref{fig:ffrt_slices}, in Figure~\ref{fig:dexm_slices} we compare ionization fields from the halo-based \texttt{DexM}-like simulations, this time showing the difference between the $R_{\rm max}$ prescription (top row) and the MFP-$\epsilon(r)$ method (second row). The effect of the MFP-$\epsilon(r)$ method on the ionization field is qualitatively similar to MFP-$\bar{\epsilon}$, i.e. the largest ionized regions shrink. The difference slices shown in the third row of Figure~\ref{fig:dexm_slices} demonstrate that the net impact is much stronger than the corresponding panels in Figure~\ref{fig:ffrt_slices}; this is mostly due to the more strongly clustered nature of the halo-based ionizing sources in \texttt{DexM} vs. FFRT. The additional small-scale clustering enhances the effect of both of our new smooth MFP prescriptions, although MFP-$\epsilon(r)$ suppresses large ionized regions somewhat more effectively than MFP-$\bar{\epsilon}$. In the bottom row of Figure~\ref{fig:dexm_slices} we show the corresponding slices using the MFP-$\epsilon$ method applied to our FFRT-P approach, demonstrating that it appears nearly identical to the more computationally involved halo-based approach.

\begin{figure*}
\begin{center}
\resizebox{16cm}{!}{\includegraphics[trim={0.7em 1em 1em 1em},clip]{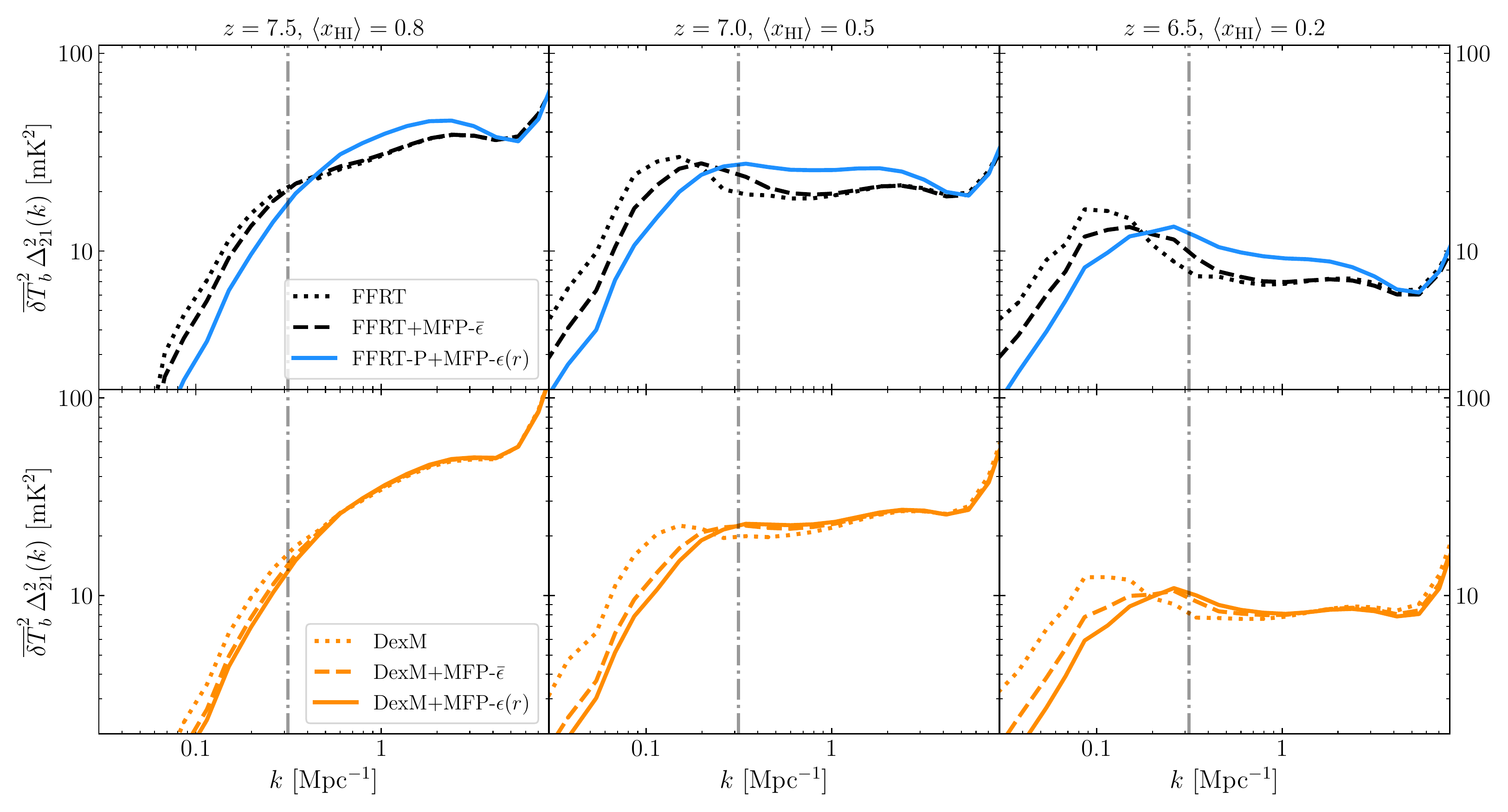}}
\end{center}
\vskip -2em
\caption{Similar to Figures~\ref{fig:sizes_redshift} and \ref{fig:pxx_redshift} but for the 21 cm power spectrum.}
\label{fig:D21_redshift}
\end{figure*}

\begin{figure*}
\begin{center}
\resizebox{16cm}{!}{\includegraphics[trim={0.7em 1em 1em 1em},clip]{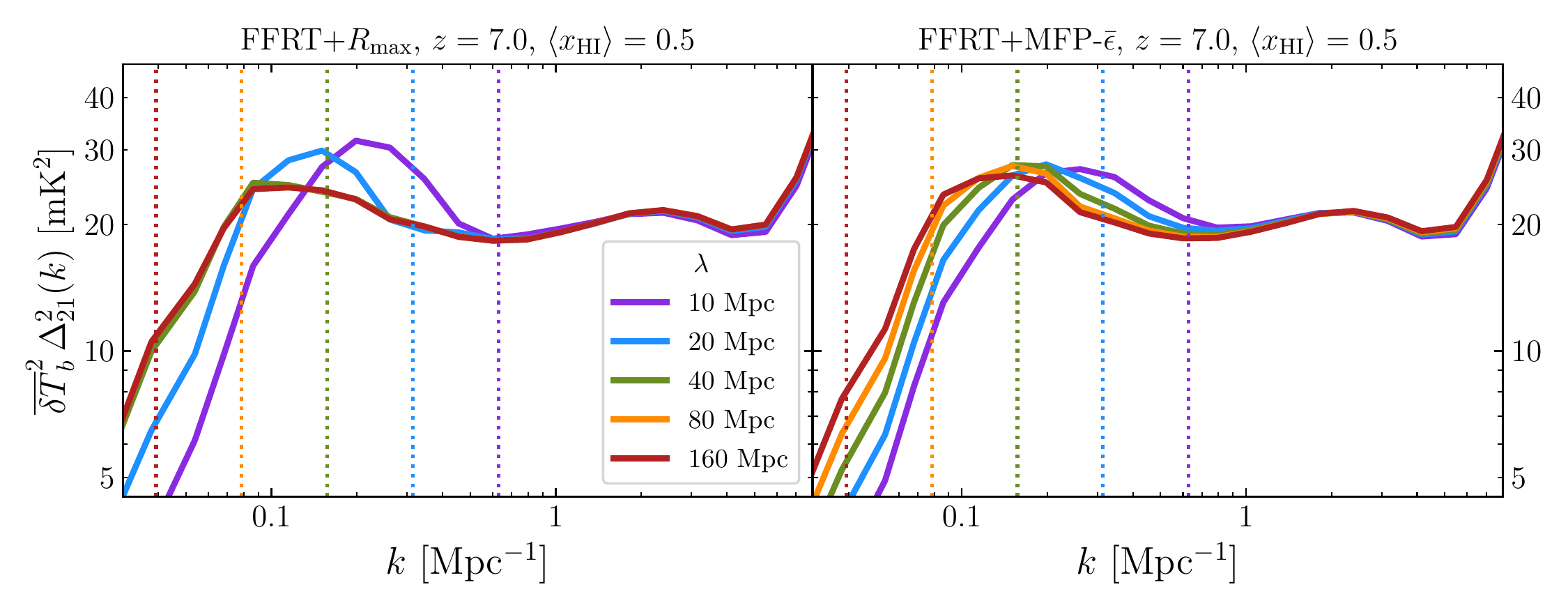}}
\end{center}
\vskip -2em
\caption{Simulated 21 cm power spectrum using FFRT as a function of $R_{\rm max}$ (left panel) and as a function of $\mfp$ using the MFP-$\bar{\epsilon}$ method (right panel). The vertical dotted lines show the corresponding $k$ values for the $\mfp$ scales shown. Note that the 40--160\,Mpc curves almost entirely overlap in the left panel, demonstrating the relative lack of sensitivity to large values of $R_{\rm max}$.}
\label{fig:D21_ffrt_mfps}
\end{figure*}

\begin{figure*}
\begin{center}
\resizebox{16cm}{!}{\includegraphics[trim={0.7em 1em 1em 1em},clip]{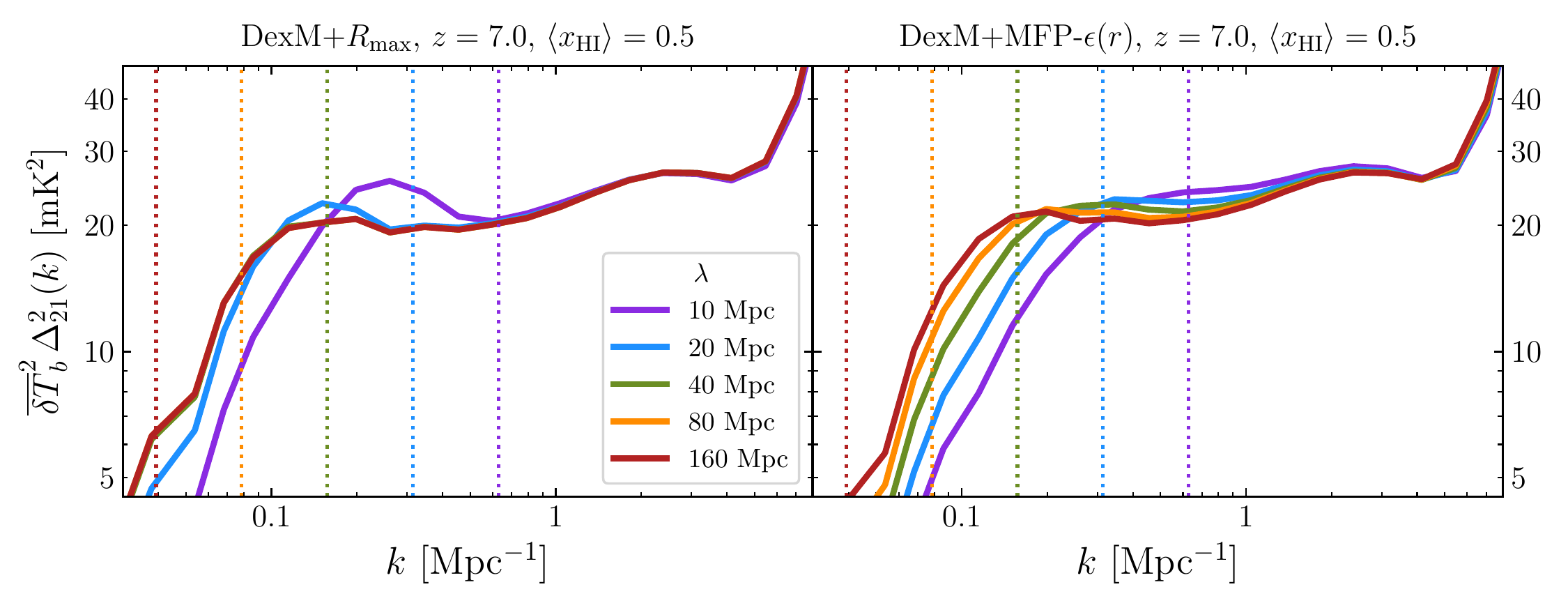}}
\end{center}
\vskip -2em
\caption{Similar to Figure~\ref{fig:D21_ffrt_mfps} but instead comparing the \texttt{DexM} method with varying $R_{\rm max}$ (left panel) to varying $\mfp$ using the MFP-$\epsilon(r)$ method (right panel).}
\label{fig:D21_dexm_mfps}
\end{figure*}

\begin{figure*}
\begin{center}
\resizebox{17cm}{!}{\includegraphics[trim={0.7em 1em 1em 1em},clip]{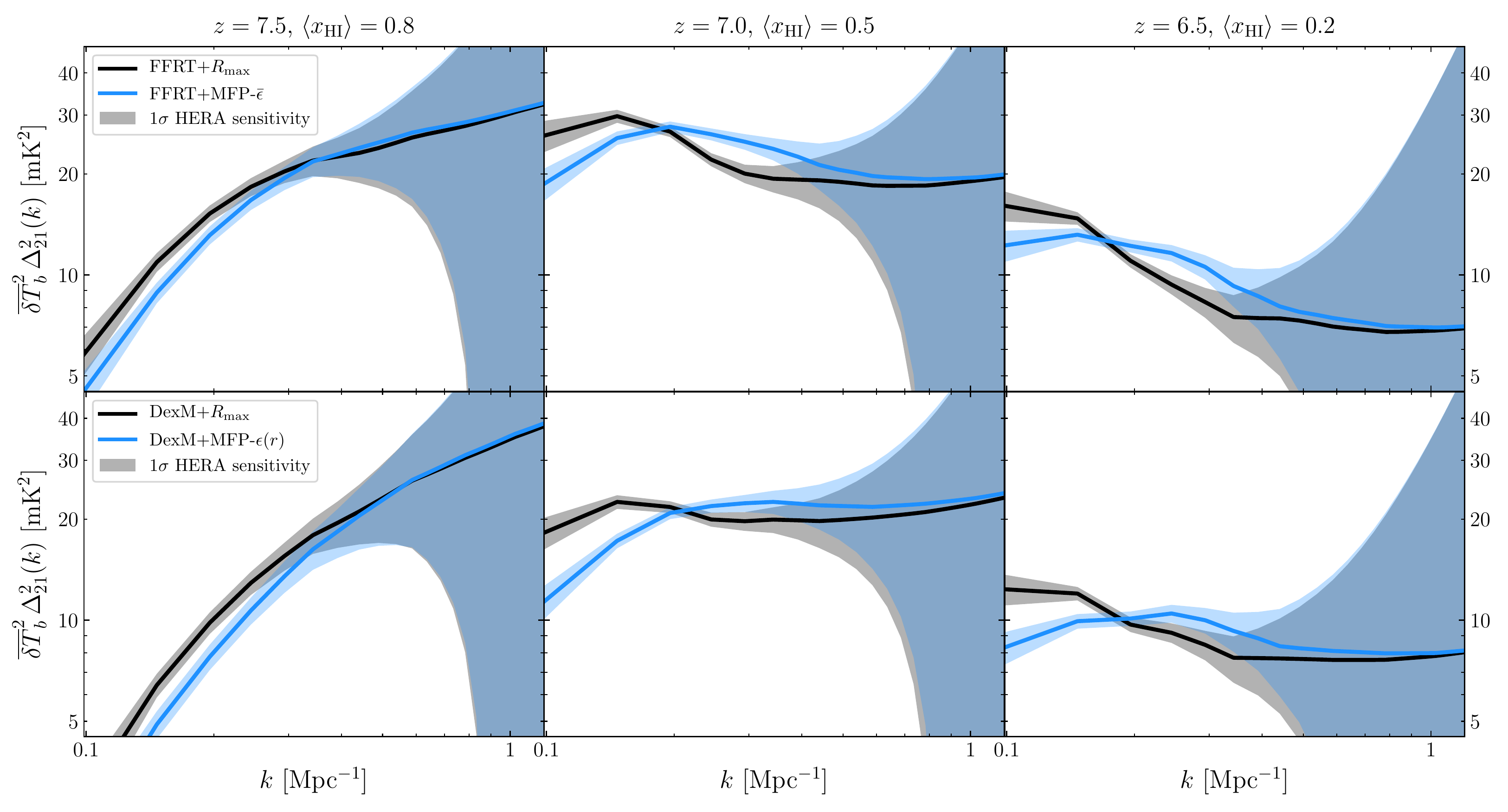}}
\end{center}
\vskip -2em
\caption{Comparison between our new mean free path prescriptions (blue) and $R_{\rm max}$ (black) in the context of the anticipated $1\sigma$ sensitivity of HERA to the 21 cm power spectrum (shaded regions). The top panels show FFRT and the MFP-$\bar{\epsilon}$ method (equation~\ref{eqn:newcrit}), while the bottom panels show \texttt{DexM} and the MFP-$\epsilon(r)$ method (equation~\ref{eqn:mfp-ebar}), all with $\mfp$ or $R_{\rm max}$\,=\,20\,Mpc. From left to right, the columns show $\langle x_{\rm HI}\rangle=0.8$, $0.5$, and $0.2$ at $z=7.5$, $7.0$ and $6.5$, respectively.}
\label{fig:D21_obs_compare}
\end{figure*}

To quantify the differences in the sizes of ionized regions between these methods, in Figure~\ref{fig:sizes_redshift} we show the ionized bubble size distributions calculated via the ``mean free path'' method\footnote{Note that this mean free path refers to the distance to the first intersection with large-scale patches of neutral gas, and not the absorption by residual neutral hydrogen in ionized regions that the mean free path refers to in the rest of the text.} from \citet{MF07}. The top panels show the FFRT methods (including FFRT-P), while the bottom panels show the \texttt{DexM} methods. As discussed above, once the ionized regions reach sizes comparable to $\mfp$, the bubble sizes are suppressed, shifting the distributions to somewhat smaller sizes. This behaviour is similar to the effect of inhomogeneous recombinations in \citet{SM14}, and similar to the radiative transfer simulations from \citet{Shukla16} which imposed additional opacity to ionizing photons from unresolved structures via an assumed $\mfp(z)$. In Figure~\ref{fig:pxx_redshift} we show the corresponding dimensionless ionization power spectra, which demonstrate the same behavior: with our new MFP implementations, power is shifted more efficiently from large to small scales, and the difference is generally larger when the spatial distribution of sources (MFP-$\epsilon(r)$) is taken into account.

\section{Implications for the 21 cm power spectrum} \label{sec:obs}

We compute the 21 cm brightness temperature assuming that the spin temperature of the gas is much higher than the CMB temperature \citep{Furlanetto06},
\begin{equation}
\delta T_b \approx 27 x_{\rm HI} (1+\delta) \left(\frac{\Omega_b h^2}{0.023}\right) \left(\frac{1+z}{10} \frac{0.15}{\Omega_m h^2}\right)^{1/2} \left(\frac{H}{dv_{||}/dr_{||}+H}\right) \rm{mK},
\end{equation}
where $\delta$ is the evolved (i.e. Zel'dovich approximation) density, $H=H(z)$ is the Hubble parameter, and $dv_{||}/dr_{||}$ is the velocity gradient along one axis of the simulation volume chosen to represent the line of sight. In Figure~\ref{fig:D21_redshift} we show the dimensionless 21 cm power spectrum $\Delta^2_{21}(k)$ multiplied by $\bar{\delta T}_b^2$, similar to Figure~\ref{fig:pxx_redshift}. The 21 cm power spectrum shows the same behavior as the ionization field, i.e. suppression of large-scale power by the new smooth mean free path prescriptions which increases towards smaller $\langle x_{\rm HI}\rangle$ as the typical ionized regions grow in size. Perhaps more interestingly, in Figure~\ref{fig:D21_ffrt_mfps} and Figure~\ref{fig:D21_dexm_mfps} we show how the 21 cm power spectrum changes with $R_{\rm max}$ or $\mfp$ for FFRT with MFP-$\bar{\epsilon}$ and \texttt{DexM} with MFP-$\epsilon(r)$ methods, respectively. Whereas adjusting $R_{\rm max}$ below 40 Mpc causes an abrupt change in the power spectrum (e.g. \citealt{Pober14}) with no variations at larger values, adjusting $\mfp$ in both new methods deforms the power spectrum more continuously. While this may imply a possible mapping between $R_{\rm max}$ and $\mfp$ to reproduce the same power spectra, the range of scales over which power is transferred is broader as well, resulting in a less prominent peak in the power spectrum for short $\mfp$ values and reducing the degeneracy between the two models.

In Figure~\ref{fig:D21_obs_compare}, we illustrate how the differences between our new mean free path methods (in blue) and the original $R_{\rm max}$ prescription (in black) compare to the sensitivity of future HERA observations. We compute the HERA power spectrum uncertainties using the \texttt{21cmSense} code \citep{Pober13,Pober14} with the default parameters (1080 hour survey, 8 MHz bandwidth, 82 channels), the ``moderate'' assumptions for foreground subtraction, and a calibration file designed to represent the HERA350 split core configuration (J. Pober, private communication). The top row in Figure~\ref{fig:D21_obs_compare} shows FFRT with and without MFP-$\bar{\epsilon}$, and the bottom row shows \texttt{DexM} with and without MFP-$\epsilon(r)$. The transfer of large-scale power to smaller scales by our new MFP prescriptions appears to be large enough to be distinguishable by HERA given the assumed observing strategy, although we note that a better (but less physical) match may be possible with a smaller $R_{\rm max}$. This points to the importance of capturing the ionizing photon absorption process accurately in the interpretation of real data.

We note that the comparison between methods in Figure~\ref{fig:D21_obs_compare} is sensitive to our choice of model parameters. In our simulations, we have assumed $M_{\rm min}=10^9$\,$M_\odot$, but sources down to the atomic cooling limit ($\sim10^8$\,$M_\odot$) may still contribute. With less biased sources, the effect of absorption would be diminished. However, we also assume that $\zeta$ is independent of halo mass, which does not reflect the expectation from abundance matching of lower star formation efficiency in low mass halos due to feedback processes (e.g. \citealt{Furlanetto17}), and act to increase the bias of sources. Finally, our fiducial mean free path of $20$\,Mpc may be too high \citep{Becker21}, which would further increase the importance of absorption.

\section{Conclusion} \label{sec:conc}

Reionization models rely on careful modeling of both sources and sinks of photons. While the former has received a great deal of recent focus (e.g. \citealt{Mutch16,Furlanetto17,Park19,Mirocha20,Hutter21}), photon absorption has received much less attention (though see, e.g., \citealt{FO05,Choudhury09,SM14,Mao20,Bianco21,Wu21Island}). However, interpretation of future observations requires accurate, physically-meaningful models for photon absorption. 
In this work we demonstrated two new methods for implementing the mean free path of ionizing photons in semi-numerical simulations of patchy reionization.  These two methods, which we call MFP-$\bar{\epsilon}$ and MFP-$\epsilon(r)$, treat the mean free path as a smooth attenuation of ionizing photons, in contrast to the sharp barrier $R_{\rm max}$ that is conventionally employed. Both methods are derived from an analogy between the excursion set reionization calculation and the integrated flux received by a point in space. While the MFP-$\bar{\epsilon}$ method can be applied to any semi-numerical reionization treatment as a simple modification of the ionization threshold, the MFP-$\epsilon(r)$ method relies on filtering the source field directly (e.g., the collapsed mass in halos), and thus should not be applied to the standard FFRT treatment in \texttt{21cmFAST}. However, we showed that a simple (but potentially ``risky'') modification of FFRT which computes a collapsed mass field via conditional Press-Schechter at the pixel scale of the simulation, which we call FFRT-P, appears to be a computationally efficient compromise that can be used with the MFP-$\epsilon(r)$ method to produce similar ionization topology as methods with explicit sources.

We demonstrated that our approach has substantial impact on 21 cm statistics, showing the importance of careful modeling of the absorption process for physical inferences from 21 cm surveys. The smooth mean free path treatments have two primary consequences for the 21 cm power spectrum. First, they generically suppress large-scale ($k\lesssim0.2$ Mpc$^{-1}$) power more effectively than $R_{\rm max}$ at a similar scale. This additional suppression comes about due to the attenuation of ionizing photons on scales \emph{below} $\mfp$, e.g. at a distance of $0.5\mfp$ one suffers from a factor of $e^{-0.5}\approx0.6$ attenuation. Second, our new methods deform the power spectrum far more smoothly and consistently with $\mfp$ compared to the sudden onset of $R_{\rm max}$ dependence once it is small enough (see, e.g., Figures~\ref{fig:D21_ffrt_mfps} and \ref{fig:D21_dexm_mfps}). The recent general adoption of the inhomogeneous recombinations prescription from \citet{SM14} in the \texttt{21cmFAST} and \texttt{21CMMC} public codes is in part due to the resulting redundancy of $R_{\rm max}$ in reionization parameter inference (see also \citealt{BP19}). In addition, we note that while we have used \texttt{21cmFAST} in this work, the fundamental idea of the approach -- incorporating uniform absorption when calculating the ionization field -- can easily be applied to other codes, such as e.g. \texttt{SimFast21} \citep{Santos10,Hassan16} or \texttt{SCRIPT} \citep{CP18,Choudhury21}.

\begin{figure}
\begin{center}
\resizebox{8.0cm}{!}{\includegraphics[trim={1em 1em 1em 1em},clip]{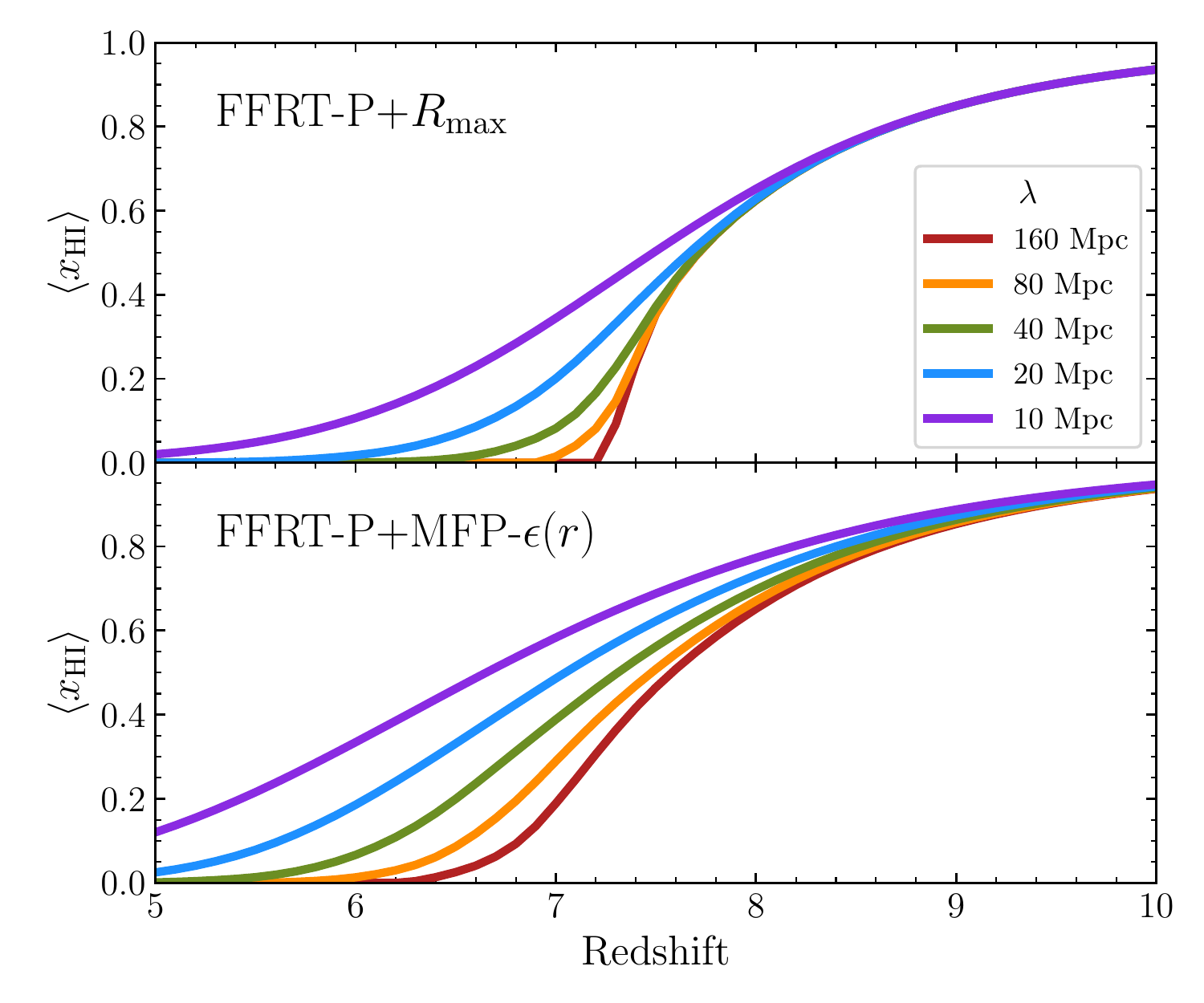}}
\end{center}
\caption{Evolution of the (volume-averaged) neutral hydrogen fraction assuming $\zeta=31.5$ and $M_{\rm min}=10^9$\,M$_\odot$, with $R_{\rm max}$ or $\mfp$ values color-coded similar to Figure~\ref{fig:D21_ffrt_mfps}. Top: FFRT-P+$R_{\rm max}$ method. Bottom: FFRT-P+MFP-$\epsilon(r)$ method.}
\label{fig:xhiz}
\end{figure}

While we have focused in this work on the effect of the mean free path on the topology of reionization as measured by the 21 cm power spectrum, the \emph{progression} of reionization should also be regulated by the same absorption. The real space top-hat filter employed here suffers from explicit non-conservation of photons, particularly late in the reionization epoch when ionized bubbles overlap (e.g. \citealt{Zahn07}), but we can still assess this impact using the methods derived in this work. We ran a suite of additional \texttt{21cmFAST} simulations from $z=12$--$5$ assuming the same $\zeta=31.5$ as the \hbox{FFRT-P+MFP-$\epsilon(r)$} $\lambda=20$ Mpc model with $\langle x_{\rm HI} \rangle=0.5$ at $z=7.0$. In Figure~\ref{fig:xhiz}, we show the resulting reionization histories for the \hbox{FFRT-P+$R_{\rm max}$} (top panel) and \hbox{FFRT-P+MFP-$\epsilon(r)$} (bottom panel) methods adopting a series of $\mfp$ values with the same colors as Figures~\ref{fig:D21_ffrt_mfps} and \ref{fig:D21_dexm_mfps}. Bearing in mind the caveat regarding photon conservation, it is clear that the MFP-$\epsilon(r)$ method produces much more extended reionization histories at fixed mean free path scale, i.e. more ionized photons are required to ionize the Universe, and is generally more sensitive to the value of $\lambda$. The mean free path, however, is measured to strongly evolve with redshift at $z\lesssim5$ \citep{Worseck14}, and may evolve even faster during the reionization epoch \citep{Park16,D'Aloisio20}, so the curves in Figure~\ref{fig:xhiz} should not necessarily be interpreted as model predictions for the true $\langle x_{\rm HI} \rangle(z)$.

Both of our new mean free path treatments require only modest additional computation time compared to the default $R_{\rm max}$ prescription in \texttt{21cmFAST}. The most significant increase\footnote{The Fourier space filter in the MFP-$\epsilon(r)$ method requires a somewhat more involved calculation (equation~\ref{eqn:mfpfilt}) than the spherical top-hat, but this adds negligible computing time in practice.} comes from the additional filter scales that must be computed between $R_{\rm max}$ and the size of the simulation box $R_{\rm box}$, 
\begin{equation} \label{eqn:overtime}
    \frac{t_{\rm MFP}}{t_{R{\rm max}}} \approx \frac{\ln{R_{\rm box}}-\ln{R_{\rm cell}}}{\ln{\mfp}-\ln{R_{\rm cell}}},
\end{equation}
where we assume that the minimum filter scale corresponds to the cell size of the simulation, $R_{\rm cell}$. This factor, $\sim$\,$1.1$--$2.1$ for the semi-numerical simulations presented in this work\footnote{As shown in Figure~\ref{fig:dxdr}, the ionization contribution from scales larger than a few times the MFP drops off very rapidly. One could then truncate the filter scales well below the entire box size in most cases; thus equation~(\ref{eqn:overtime}) is a conservative overestimate to the required extra time.}, is small compared to the additional complexity of fully numerical radiative transfer approaches. 

We note that the methods described in this work have assumed a spatially-uniform mean free path, but this may not be an accurate approximation in the late stages of reionization due to the interplay with the fluctuating ionizing background \citep{McQuinn11,Crociani11,DF16} or variations in the degree of post-reionization dynamical relaxation depending on the local reionization redshift \citep{D'Aloisio20,Cain21}. While the MFP-$\epsilon(r)$ method is not well suited to such fluctuations because the mean free path is built in to the excursion set filter, the MFP-$\bar{\epsilon}$ approach could in principle be employed in a manner similar to \citet{SM14} with varying $\mfp$ depending on the local physical state (e.g., density and ionizing background intensity). We will explore implementing a fluctuating mean free path into our new absorption prescriptions in future work. 

Following the submission of this work, \citet{Becker21} performed the first direct measurement of the opacity of the IGM to ionizing photons at $z\sim6$ along the line of sight to luminous quasars. After correcting for the over-ionization of gas close to the quasars, they estimate a mean free path in the general IGM of $\mfp = 0.75^{+0.65}_{-0.45}$ proper Mpc (although the lower bound may be somewhat higher, see \citealt{Bosman21MFP}). This mean free path is much shorter than predicted by an extrapolation from measurements at lower redshift. Such a short mean free path implies that the reionization process was likely strongly regulated by absorption \citep{Cain21,Davies21}.

In the next decade, we expect that 21 cm experiments and other probes of the reionization era will begin to pin down the history and morphology of that process. Here we have shown that absorption should have a significant effect on the ionization field and presented efficient techniques to model it accurately. We hope that these and similar physical models will yield improved constraints from future observations.

\section*{Acknowledgements}

We thank Jonathan Pober for sharing his HERA configuration files for \texttt{21cmSense}, Andrei Mesinger for useful discussions, and the ENIGMA group at UCSB/Leiden for helpful comments. SRF was supported by the National Science Foundation through award AST-1812458. In addition, this work was directly supported by the NASA Solar System Exploration Research Virtual Institute cooperative agreement number 80ARC017M0006. This material is based upon work supported by the National Science Foundation under Grant Nos. 1636646 and 1836019 and institutional support from the HERA collaboration partners.  This research is funded in part by the Gordon and Betty Moore Foundation.

\section*{Data Availability}

The data underlying this article will be shared on reasonable request to the corresponding author.

\bibliographystyle{mnras}

 \newcommand{\noop}[1]{}

\appendix

\section{Bridging the gap between FFRT and \texttt{DexM}}

\begin{figure*}
\begin{center}
\resizebox{16.0cm}{!}{\includegraphics[trim={0 0 0 0},clip]{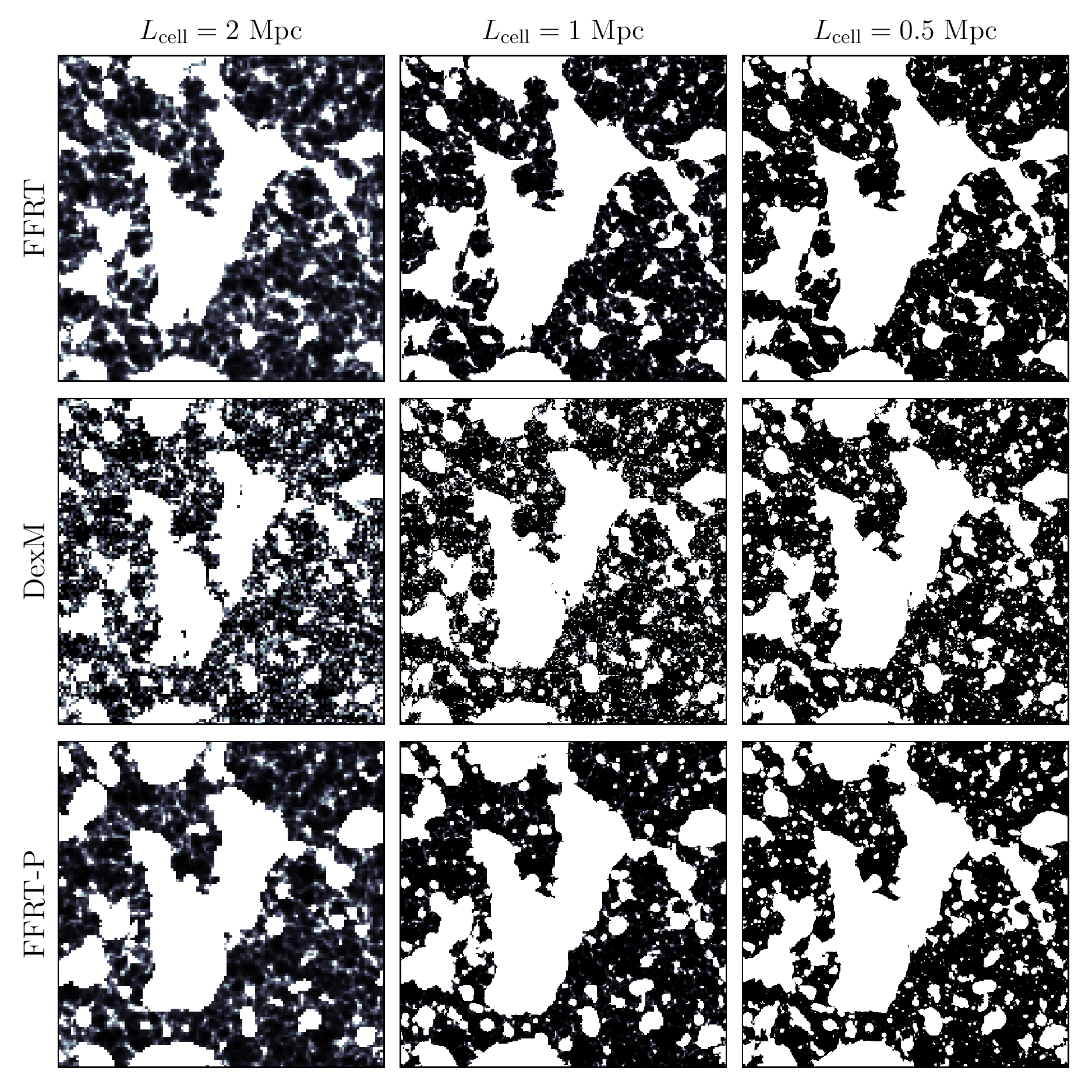}}
\end{center}
\vskip -2em
\caption{Ionization field slices using FFRT (top row), \texttt{DexM} (middle row), and FFRT-P (bottom row) with varying spatial resolution. From left to right, the ionization fields were computed on grids of $128^3$, $256^3$, and $512^3$ cells, corresponding to cell sizes of $L_{\rm cell}=2.0$, 1.0, and 0.5 Mpc, respectively.}
\label{fig:slice_res}
\end{figure*}

\begin{figure*}
\begin{center}
\resizebox{16.0cm}{!}{\includegraphics[trim={0 0 0 0},clip]{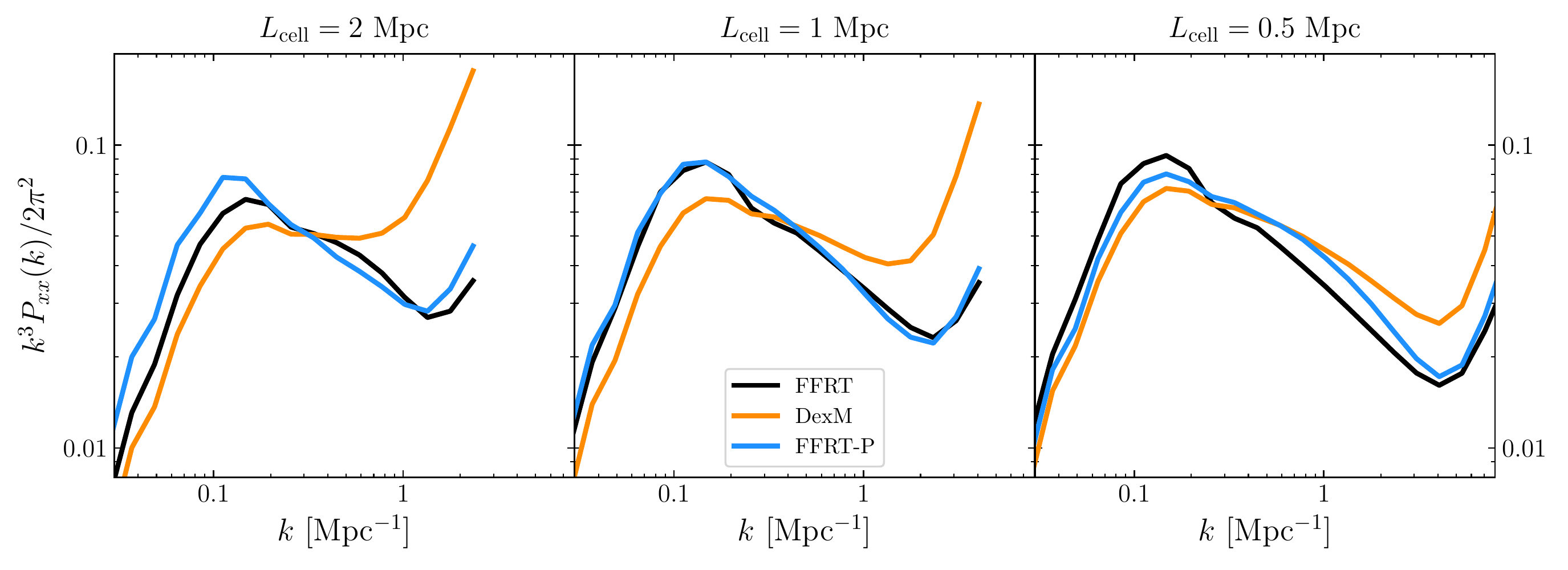}}
\end{center}
\vskip -2em
\caption{Power spectra of the ionization fields shown in Figure~\ref{fig:slice_res}.}
\label{fig:pxx_res}
\end{figure*}

Unlike the standard FFRT method, the MFP-$\epsilon(r)$ approach described in Section~\ref{sec:mfp} can only be applied to fields of collapsed mass, e.g. from a halo catalog, but this is potentially computationally cumbersome. In this Appendix we describe a method that computes the collapsed mass field directly from the density field, and which retains the speed and simplicity of the FFRT algorithm while producing ionization fields more consistent with discrete source models.

The standard implementation of FFRT computes $f_{\rm coll}$ at each filter scale from the smoothed density field following a conditional Press-Schechter scheme \citep{LC93}. However, as remarked in footnote 4 of \citet{SM14}, it is theoretically possible to compute $f_{\rm coll}$ at the pixel scale of the simulation and instead filter the collapsed \emph{mass} field directly, similar to discrete source models, although at the risk of spurious resolution-dependent features. In \texttt{21cmFAST}, $f_{\rm coll}$ is already computed by applying a prescription meant for the linear density field to a quasi-non-linear density field computed via the Zel'dovich approximation, so the $f_{\rm coll}$ on the smallest (i.e. most non-linear) scales may be quite dubious. In the following we explore this ``risky'' method, which we call ``FFRT-P,'' and find that (in certain regimes of spatial resolution) it can actually reproduce the ionization field morphology of discrete source models more accurately than FFRT.

To compare the simulation methods, we adopt the same semi-numerical simulation box as in the main text, with $4096^3$ initial conditions in a volume $256$ Mpc on a side, and generate evolved density and ionization fields at $z=7.0$ with $\langle x_{\rm HI}\rangle=0.5$ on grids of $128^3$, $256^3$, and $512^3$, corresponding to cell sizes of 2, 1, and 0.5 Mpc, respectively. For the purposes of this comparison we assume the standard $R_{\rm max}$ prescription for absorption for simplicity, but note that the results are largely unchanged with the new MFP methods described in the main text. Figure~\ref{fig:slice_res} shows slices of the resulting ionization fields, and Figure~\ref{fig:pxx_res} compares the ionization field power spectra. While FFRT-P overestimates the large-scale power relative to FFRT and \texttt{DexM} when cell sizes are large, for cell sizes below $\sim1$ Mpc it appears to converge towards \texttt{DexM}. At cell size $L_{\rm cell}=0.5$ Mpc the agreement between FFRT-P and \texttt{DexM} is very good except for $k>1$ Mpc$^{-1}$, where the small-scale power in FFRT-P is suppressed due to a lack of shot noise from discrete halos. Interestingly, while the FFRT and \texttt{DexM} ionization power spectra change considerably on large scales as a function of resolution (see also \citealt{CP18}), the FFRT-P ionization power spectrum is much more stable. Indeed, the apparent excess in large-scale power of FFRT-P at low resolution is much more consistent with the other methods at higher resolution.

One potential limitation of FFRT-P is that, due to the application of the same collapse fraction prescription as FFRT, the simulation cells must be large enough that halos more massive than the minimum halo mass can form out of them, i.e. $M_{\rm cell} > M_{\rm min}$. In FFRT this is not necessary because the minimum filtering scale is a free parameter, although that limits the small-scale structure in the resulting ionization field.

We note that \citet{Trac21} have more thoroughly tested similar pixel-scale excursion set methods which they dub ``ESF-L'' and ``ESF-E,'' where FFRT-P is most similar to their ESF-E method. In general, the ESF-L method appears to provide a substantially more accurate representation of the collapsed mass field than ESF-E (and thus, FFRT-P), at no additional computational expense. In future work, we will implement ESF-L in our semi-numerical simulations.

\bsp	
\label{lastpage}
\end{document}